\documentclass[12pt]{article}
\usepackage[super,sort&compress,comma,numbers]{natbib} 
\usepackage[version=3]{mhchem}
\usepackage{graphicx} 
\usepackage{lastpage}
\usepackage{subcaption}
\usepackage[format=plain,justification=justified,singlelinecheck=false,labelfont=bf,labelsep=space]{caption}
\usepackage{float}
\usepackage{fancyhdr}
\usepackage{url}
\usepackage{array}
\bibpunct{}{}{,}{s}{,}{,}
\usepackage{natmove}
\usepackage{comment}
\usepackage[usenames,dvipsnames]{xcolor}
\newcommand\blfootnote[1]{%
  \begingroup
  \renewcommand\thefootnote{}\footnote{#1}%
  \addtocounter{footnote}{-1}%
  \endgroup
}

\usepackage{epstopdf}

\definecolor{cream}{RGB}{222,217,201}

\title{Classifying Antimicrobial and Multifunctional Peptides with Bayesian Network Models}
\author{Rainier Barrett\textit{$^{b}$}, Shaoyi Jiang\textit{$^{a*}$}, Andrew D White\textit{$^{b*}$}}

\begin{document}

\newcommand{\SI}{Supporting Information}

\maketitle 
\blfootnote{\textit{$^{a}$~University of Washington, Department of Chemical Engineering, Seattle, WA, USA. Tel: 01 206 616 6509; E-mail: sjiang@uw.edu}}
\blfootnote{\textit{$^{b}$~University of Rochester, Department of Chemical Engineering, Rochester, NY, USA. Tel: 01 585 276 7395. E-mail: andrew.white@rochester.edu }}

\begin{abstract}
Bayesian network models are finding success in characterizing enzyme-catalyzed reactions, slow conformational changes, predicting enzyme inhibition, and genomics. In this work, we apply them to {\color{black}statistical} modeling of peptides by simultaneously
identifying amino acid sequence motifs and {\color{black}using a motif-based model to clarify the role motifs may play in antimicrobial activity}. We construct models of increasing sophistication, demonstrating how
chemical knowledge of a peptide system may be embedded  without requiring new derivation of model fitting equations after changing model structure. These models are
used to construct classifiers with good performance {\color{black}($94\%$ accuracy, Matthews correlation coefficient of 0.87)} at predicting
antimicrobial activity in peptides, while at the same time being built of interpretable parameters. We demonstrate use of these models to
identify peptides that are potentially both antimicrobial and antifouling, and {\color{black}show that the background distribution of amino acids could play a greater role in activity than sequence motifs do}.
This provides an advancement in the type of peptide activity modeling 
that can be done and the ease in which models can be constructed.
\end{abstract}

\section{Introduction}

{\color{black}Bayesian networks are a statistical modeling framework that are ideally suited for use in combination with a quantitative structure-property relationship (QSPR) modeling framework due to their ability
to encode chemical knowledge and design interpretable models.\cite{GhahramaniReview,QSPRBayes} QSPR modeling (alternatively known as QSAR for ``Quantitative Structure-Activity Relationship'') is a term used to refer to a suite of statistical modeling techniques. Its purpose is not always specifically to identify structural features and link them to activity, but to identify and/or exploit trends among the features of a chemical dataset in order to make statistical predictions. This broad class of modeling methods makes use of input data such as chemical descriptors and peptide sequences. Recent reviews may be found in \citet{QSARReview, Cherkasov2014}, and \citet{Nongonierma2016}}. The power of Bayesian network
models lies in their ability to treat sophisticated models 
with general training techniques. Due to the
generality of the training algorithms, even datasets like the massive ENCODE database can be analyzed.\cite{encodenature2012} A recent review of them may be
found in Ghahramani.\cite{GhahramaniReview} Recent examples of applied Bayesian networks include predicting microcanonical melting points,\cite{BayesianMeltingPoints} modeling a terrorist network,\cite{BayesianTerrorism} and extrapolating clinical trial results beyond the original demographic.\cite{BayesianClinicalTrials}

When applying a Bayesian network model to small drug-like molecules,
one could specify that the molecule must have a molecular weight below
a cut-off, and that at least two QSPR descriptors must be in a certain range. Such
constraints are difficult to embed into linear discriminant analysis,
for example, and require a new derivation of the model fitting
procedure. There is no such requirement in Bayesian networks, due to the
generality of their training procedures. This generality also means
that fast algorithms have been developed that make use of such
constraints to reduce the training space. The combination of the constraints and
speed allows models to be constructed that are easily
interpretable.\cite{membraneTopology} 
In particular, Bayesian models have recently been employed in {\color{black}specifically biological} studies of a wide range of important topics, including enzyme-catalyzed reactions,\cite{ChouEnzymesGraphTheory,ChouExampleRateLaws,ChouExampleEnzymes,ChouExampleKinetics} slow conformational change,\cite{GrpahicalSlowConformationalChange} and inhibition of HIV-1 reverse transcriptase.\cite{GraphicalHIVInhibition}


Another benefit of Bayesian networks is their ability to do
multimodal modeling.\cite{integrativeModel} Multimodal modeling
is the combination of multimodal data into a unified model.\cite{multimodal} For
example, combining the sequence data and chemical
descriptors of a peptide is a challenging task. In a Bayesian network,
two parts of the model may deal with the different data types and be
connected through probability distributions. This has an advantage
over other model combination techniques, such as consensus
modeling,\cite{qsarConsensus} in that both models may be
trained independently and later combined, instead of a step-by-step procedure. Applying
these types of models to chemistry problems will open new ways of combining
data such as bioavailability descriptors, sequence models, and
simulation results.

In recent years, a number of studies of antimicrobial peptides (AMPs) have been made using machine learning techniques.
The Antimicrobial Peptide Database (APD)\cite{apd} was designed to collect known peptides with antimicrobial properties. Past studies involving AMP classification include: \citet{amper} who created AMPer using a hidden Markov model approach,  \citet{antibp} who developed AntiBP (later improved by \citet{antibp2}) to classify antimicrobial peptides using sequence information, and \citet{CAMP} who created the Collection of Anti-Microbial Peptides, a database of AMPs with built-in tools for prediction and analysis. Finally, \citet{AMPClassifier} used machine learning techniques to classify AMPs by target (bacteria, viruses, etc.) in addition to anti-microbial activity alone.

However, antimicrobial activity \textit{in vitro} is not necessarily indicative of broad applicability for other uses. In complex media, fouling (non-specific binding at the surface of a material) leads to a loss of activity.\cite{WhyAntifouling2} Thus, in order for an AMP to be of use in applications such as
biomedical devices and marine
coatings,\cite{WhyAntiFouling1,WhyAntifouling2,WhyAntifouling3} it is necessary to ensure that AMP retains both antimicrobial and antifouling properties in complex media. To this end, we have constructed accurate
 Bayesian 
 models that can predict antimicrobial activity, identify
sequence motifs, elucidate important descriptors, and identify potential
multi-functional peptides that are both antimicrobial and antifouling.

In this work, Bayesian network models are created to predict antimicrobial
activity and identify possible multifunctional peptides. Two datasets are
used to train the networks. The first is 351 unique peptides
which inhibit growth of gram-positive bacteria from the APD.\cite{apd} The second is a collection of
approximately $3,600$ sequence fragments from the surface of human proteins. The second
dataset is hypothesized to contain sequences which resist nonspecific
interactions in biological systems.\cite{White2012Decoding, White2012Role} Bayesian models are well-suited to small datasets such as these, as they have demonstrable ability to be accurately trained on datasets with as few as $100$ points. \cite{SmallDatasetBayesNets}

In the materials and methods section, we describe the construction of the datasets, the training procedure used for the models, and the descriptors used. In the
results section, we  examine model accuracy, compare our models with a simpler traditional 
 machine learning approach, and finally,
identify two peptides that are predicted to have antimicrobial and
antifouling properties.

\section{Materials and Methods}

\begin{figure}

\centering
\includegraphics[width=3.5 in]{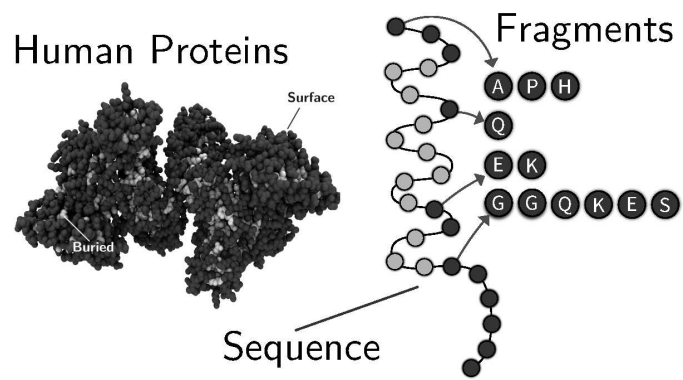}
\caption[Overview of human protein database construction]{Diverse
  proteins isolated from humans with structure in the protein data
  bank created a database of 1,162 proteins. The surface was found as
  described in \citet{White2012Decoding}. Contiguous surface sequences
  were found (dark gray) and converted into sequence fragments. All
  with length greater than 4 were used.}
\label{fig:qspr2:cartoon}

\end{figure}

It is critical in the development of a sequence-based statistical prediction method for biological systems that the process be clear and easily reproducible. As suggested by \citet{Chou5Steps} and demonstrated in a series of recent articles from that research group,\cite{ChouExampleRateLaws, ChouExampleEnzymes, ChouExampleKinetics, AMPClassifier} a predictive statistical model for biological systems should use the following guidelines to achieve its goals: construct benchmark datasets for training and testing the predictor, create a suitable mathematical expression that represents the relevant properties for prediction, implement an appropriate model and training algorithm, perform cross-validation tests to evaluate the accuracy of prediction, and establish a web-based public interface for ease of use. In the following, we address each of these steps in turn.

{\color{black}In order to build a classification model to identify peptides that could be both antimicrobial and antifouling,} two datasets were used. The first is from the APD as of September 2017\cite{apd} and contains 482 sequences
which show activity against gram-positive bacteria. The 482 sequences
were reduced to 351 by removing similar sequences. Here, ``similar'' sequences
were defined as those separated by 2 or fewer single position
substitutions (e.g, EDGRT and ADGRS are similar). 
This definition of sequence similarity has no inherent chemical meaning. It was chosen as a method of
removing certain sequences to reduce over-representation of combinatorial studies,
where single positions are changed over multiple trials. Although one substitution can be enough to drastically change the activity of a peptide,\cite{SingleSubstitution3, SingleSubstitution2, SingleSubstitution1} it is still necessary to avoid including these similar sequences due to the statistical nature of the model. Even if changing one or two amino acids affects the activity, including many peptides with nearly identical sequences would bias the model toward the (potentially very long) unchanged parts of these peptides.
The second dataset, ``Human'', is
built upon the protein dataset from \citet{White2012Decoding} All
\textit{contiguous} amino acid sequences of \textit{length greater than 4} 
present on the \textit{surface} of proteins from that dataset were tabulated as
independent sequences, as depicted in Figure
\ref{fig:qspr2:cartoon}. This yields 3,600 unique sequences.

Another aspect of antimicrobial activity that our datasets do not address is post-translational modification (PTM) of peptides. In fact, $1147$ out of $1755$ of the peptides in the APD database are known to undergo PTM before activity.\cite{PTMinAPD} However, the goal of this model is not to explain all factors that lead to antimicrobial activity, but to accurately predict potential antimicrobial activity with as little information as possible, i.e. \textit{only} sequence and/or chemical descriptor information.

To be of interest for applications like screening and other biochemical experiments, it is crucial to minimize the false positive rate (FPR) of the predictive model. Reducing false positives prevents 
 the waste of experimental time and resources by precluding the investigation of an incorrectly-predicted candidate peptide.
Furthermore, the ability to reject false positives is an important attribute of the
 model itself, because it indicates that the model is not over-fit. 
However, to evaluate performance in this regard, negative data must be either gathered or generated; no one has tabulated a list of peptides which are {\it not}
antimicrobial. \citet{ann1} approached this problem by using sequences
not reported to have activity, which may be a good assumption since
AMPs are likely rare. We use the same approach
here to evaluate model performance. A decoy dataset was generated by
replacing each residue in the APD dataset with a randomly selected
amino acid drawn from the distribution of amino acids among all entries in the Protein Data Bank.\cite{PDB}

\begin{figure*}
	
	\centering
	
	\includegraphics[width=\linewidth]{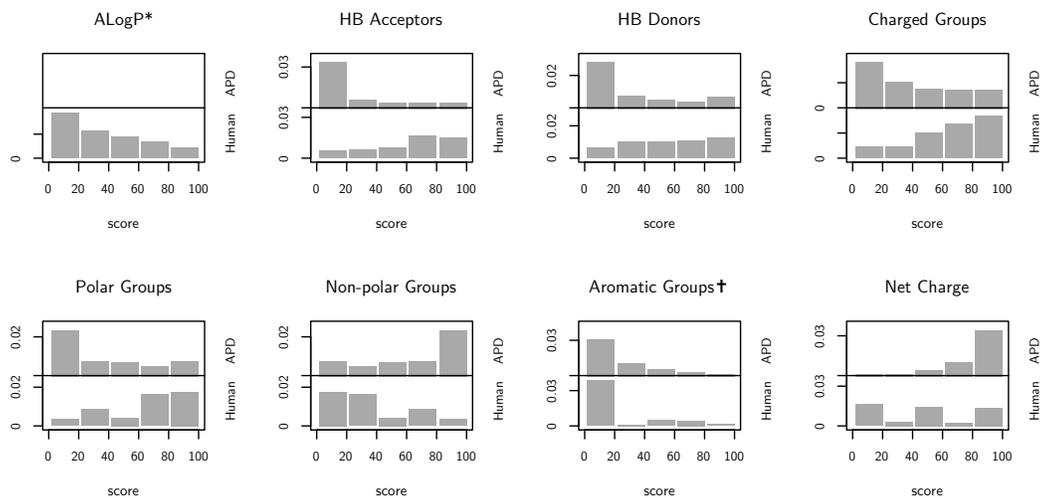}
	\caption{Histogram plots of descriptors of the {\color{black}Antimicrobial Peptide Database (APD)} and Human
		datasets. {\color{black}``HB'' stands for hydrogen bond.} A ranking of 100 indicates a high value of the descriptor
		relative to all peptides. $^*$ALogP is not calculated for the APD due to its
		poor accuracy at long peptide lengths found in the APD
		dataset. $^\dagger$ The Human aromatic histogram is skewed because
		most sequences in the ``Human'' dataset have no aromatic groups because they are drawn only from the surface of human proteins.}
	\label{fig:qspr2:descriptors}
	
\end{figure*}

The following descriptors were considered: ALogP, number of hydrogen bond acceptors, number of hydrogen bond donors, number of charged groups, number of polar goups, number of non-polar groups, number of aromatic groups, and net charge. {\color{black}We investigated these descriptors with the goal of finding data that would distinguish our two datasets from one another and from the space of all peptides in general.} All descriptors, except ALogP, were calculated using the peplib R plugin.\cite{White2013Standardizing} 
ALogP was calculated according to \citet{Ghose1987ALogP} as implemented in the Chemistry Development Kit.\cite{CDK} To estimate the distribution of ALogP values, the
calculation was performed on all possible amino acid sequences of lengths from 1 to 3, and on a
random sample of 5,000 peptides, drawn uniformly from the datasets found in \citet{Sweeney2005SHP} and \citet{Chen2010TULA} for each length up to 10.
 Using a procedure described in the \SI, chemical descriptors are first
converted into rankings that span 0 to 100. The rankings represent the value of the descriptor relative to all possible peptides in the dataset. Except for the ALogP descriptor, the descriptors used in this work are all additive, i.e. they are cumulative sums of the descriptors of all individual amino acids in a given peptide. {\color{black}The results of these calculations were tabulated as a single csv file for each dataset, each line of which consists of a peptide (specified as a single-letter amino acid sequence) paired with the list of chemical descriptors calculated for that sequence. In the chemical descriptor model, only the descriptor values are used for training. The motif model uses only the sequences themselves as input data. Withheld testing data was selected as a random 20\% of input data for both models.}

The descriptor rankings were calculated on the two datasets and are
shown in Figure \ref{fig:qspr2:descriptors}. Each box shows the
histogram of the descriptor rankings. A flat histogram indicates the
ranks are distributed identically to all possible peptides and thus
the descriptor is likely unrelated to the activity. As was desired,
all the descriptors chosen distinguish both datasets from all possible
peptides and one another. The APD dataset shows an abnormally low
number of charged residues and a high number of non-polar
groups. Consistent with past analysis of AMPs,\cite{structure, ReadDiscussionDontJustCite, deNovoAMP} the APD dataset has a lower number of charged residues with a skew toward positively charged.  The number of
charged residues is high on human protein surfaces, as seen
previously.\cite{White2012Decoding} This is also reflected in the high
water solubility (low ALogP values). There is no dominant net charge
in one direction or another for human protein surfaces. The number of
aromatic residues is low, which is expected, since the number of
aromatic residues is low across all proteins generally, and hydrophobic aromatic side chains mostly occur on the interior of human proteins.

\subsection{Model Description}

\begin{figure}
\centering

\includegraphics[width=8cm]{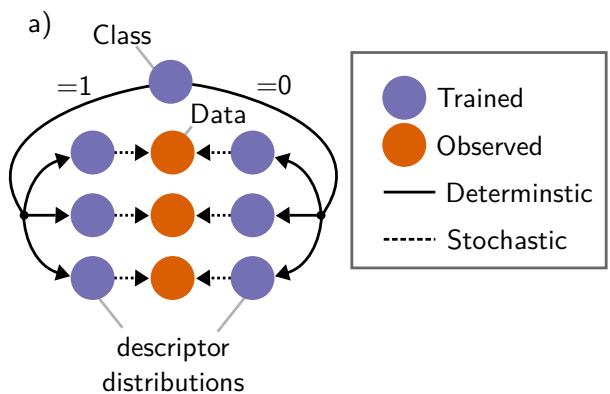}
\caption[2-state classifier graph and results]{A graphical representation of a 2-state
  classifier that fits 3 observed descriptors to either distribution 0
  or 1. A class of 1 indicates activity. In this work, the ``Trained'' nodes are our mixture-of-Gaussians distributions, and the three ``Observed'' nodes correspond to the three chemical descriptors chosen.}
\label{fig:qspr2:classifier}

\end{figure}

The chemical descriptor model (henceforth the ``QSPR model'')  was treated as a one-dimensional, two-state Gaussian mixture with respect to each descriptor. In Gaussian mixture modeling, the underlying distribution of a set of observations is estimated by fitting a function made up of a sum of $k$ Gaussian kernels. The hyperparameters are the means, heights, and variance matrices, which are fitted for each kernel. Some background on this technique can be found in \citet{GaussmixImageProcessing} and McNicholas and Murphy.\cite{ParsimoniousGaussmix}
This model is one-dimensional in that each chemical descriptor was fitted to its own separate mixture-of-Gaussians distribution, with no correlation between distributions. The number of Gaussian kernels was varied from 1 to 10, with the best performance given by the 3-kernel set (see discussion and \SI). This portion of the model was implemented using the PyMC3 package for Python 3.\cite{pymc3} This two--state mixture model was used to classify sequences as
antimicrobial using the descriptors in Figure
\ref{fig:qspr2:descriptors}, and is shown in graphical representation in
Figure \ref{fig:qspr2:classifier}. Three descriptors were chosen based on inspection of
Figure \ref{fig:qspr2:descriptors}: net charge, number of non-polar
groups, and number of charged groups{\color{black}, as these had the most distinct histograms between the two datasets used.}

Next, a motif model was constructed which classifies sequences based
on the existence of motifs in the sequence. Emphasis was
placed on keeping the model interpretable to non-experts.
For this reason the following attributes were chosen: 
($1$) There are $0$ to $k$ possible motif classes that may be observed. 
($2$) Each peptide belongs to only one motif class.  
($3$) Motifs may not be partially expressed. 
($4$) Non-motif residues are drawn from a ``background'' distribution that is shared among all peptides in the dataset. 
($5$) Motifs are of fixed length $w$, and motif distributions are sparse, i.e. usually only one amino acid exists in each position. 
($6$) The probability of a motif starting at a sequence position is independent
of the position.
{\color{black}($7$) Motif distributions should be sparse, i.e. they should have few non-negligible entries. } This model formulation is in part inspired by previous work in motif identification, such as \citet{Bailey1995Unsupervised} and Schwartz and Gygi.\cite{SchwartzMotifDiscovery2005}
The features that are unique to this description relative to past
motif models \cite{Bailey1994MEME, White2013Standardizing} are the
regularization of motifs, a tied background distribution, independent
motif start probability, and the ability to deal with variable length
sequences. The regularization {\color{black}method chosen (L1 regularization, see Section 2.2)} forces the motifs to be sparse so that
each motif position only has one or two possible residues. This makes
motif interpretation more intuitive. The tied background distribution
reduces the number of model parameters by $(k - 1)(A - 1)$, where $A$
is the number of amino acids. {\color{black}This background distribution can also be used as the motif model by itself if we let $k=0$. This ``Background-only'' model is a limiting case of the motif model, but is otherwise specified and trained in an identical fashion.} Such a change greatly complicates
traditional algebraic analysis of the model, but is simple to include with this formulation. The uniform motif start probability reduces the
number of parameters by $k(l - 1)$, where $l$ is the length of the
sequences. The ability to deal with variable length sequences without
pre-alignment is a significant feature and is what allows modeling of
the highly heterogeneous APD. The model description is thus-far
complex, but the trade-off is that the parameters that are derived
from this model are intuitive and few.
The motif model was implemented directly in Python as an extension module written in C++. 
 Complete model specifications can be found in the \SI.

\begin{figure}

\centering
\includegraphics[width=3.5 in]{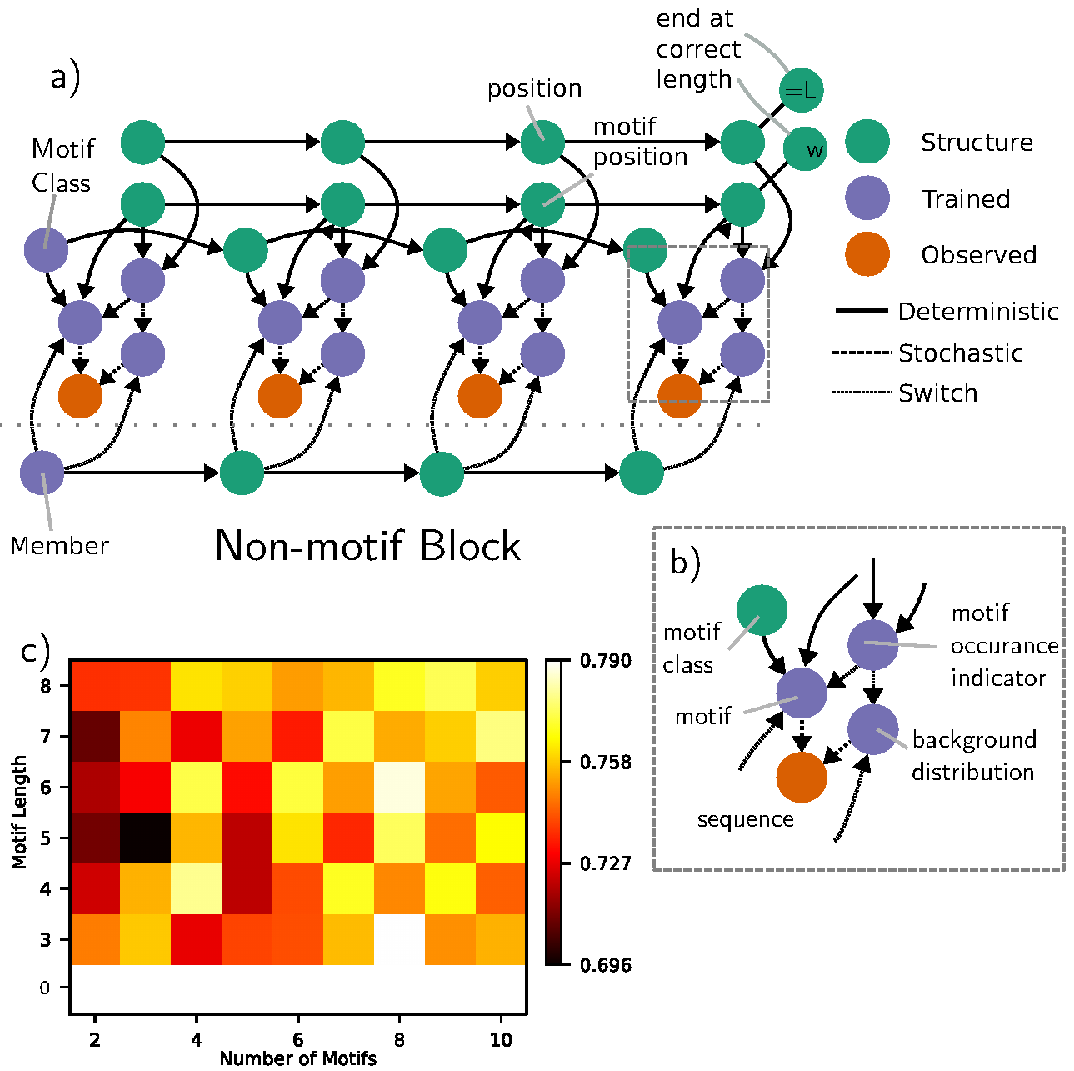}
\caption[Graph of motif model and results on motif width and number
  choice]{Panel a is the graphical representation of the motif model used. The middle
  parts of the graph are repeated as many times as necessary to
  fit the length of a sequence (length $4$ depicted). Panel b is an inset showing how the
  motif indicator governs whether the probability for a given amino acid is drawn from the background or motif distribution. Panel c is the prediction
  accuracy as a function of the motif width and motif number. A motif length of $0$ indicates the performance of the background-only model. The maximum prediction accuracy ($79\%$) of the motif model was the same as the background-only model.}
\label{fig:qspr2:dgm}

\end{figure}

\begin{figure}[h]
    \centering
    \begin{subfigure}[b]{0.45\linewidth}
    \centering
    \includegraphics[width=3 in]{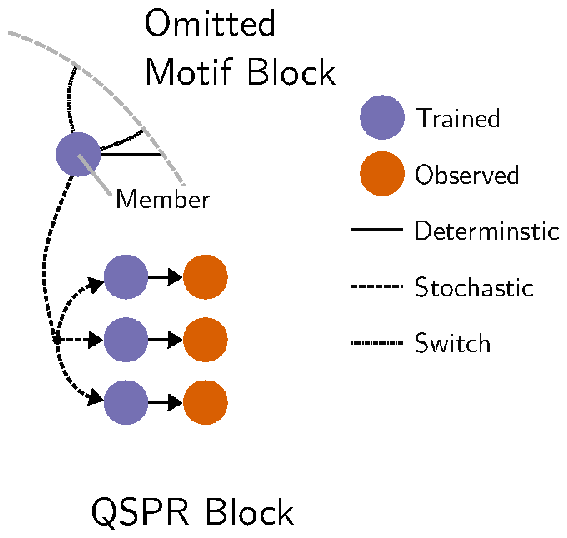}

    \subcaption[Graph of combined QSPR/Motif model]{Graph of combined
    QSPR/Motif model. The ``omitted motif block'' refers to
    Figure \ref{fig:qspr2:dgm}a. The ``Member'' value drawn from the model shown in Figure \ref{fig:qspr2:classifier} is used as the membership value in the rest of the model (Figure \ref{fig:qspr2:dgm}a.)}
    \end{subfigure}
    ~
    \begin{subfigure}[b]{0.45\linewidth}
    \centering
    \includegraphics[width = 3 in]{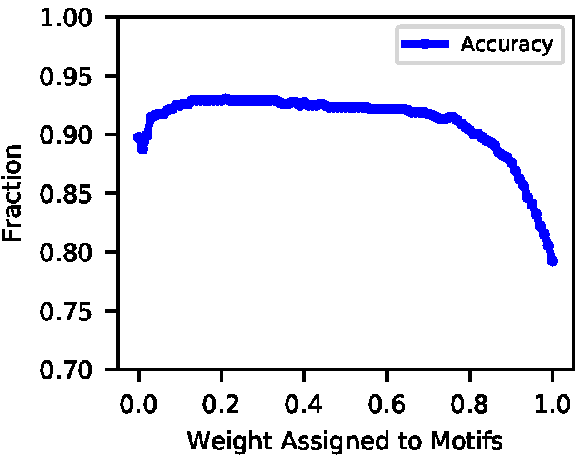}
    \subcaption{Performance of the combined QSPR+motif model on the APD dataset with distributions from the 3-kernel QSPR model and the {\color{black}8-motifs, length-3} motif model. The x-axis is the weight assigned to the Motif half of the model ($0$ is no influence and $1$ is total influence).}
    \end{subfigure}

    \caption{Graphical representation and classification performance of the combined QSPR/motif model.}
    \label{fig:combinedmodel}

\end{figure}

The last classifier considered combines features from the previous
two. Its graphical representation is shown in Figure~\ref{fig:combinedmodel}a. In this model, the ``Class'' node shown in Figure~\ref{fig:qspr2:classifier} is connected to the ``Member'' node shown in Figure~\ref{fig:qspr2:dgm}a. This indicates
incorporation of the descriptor distributions shown in Figure
\ref{fig:qspr2:classifier} into a new ``QSPR'' block in the graph in Figure \ref{fig:qspr2:dgm}a. Thus, the model takes in both sequence information and descriptors. The
best-performing motif number ($k=8$) and motif width ($w=3$) from the motif model, and the same
descriptors (net charge, number of charged groups, and number of
non-polar groups) from the QSPR model were used. The complete model specification is given in the
\SI. 

\newpage
\subsection{Model Training and Validation}

The QSPR model was trained using the built-in Metropolis-Hastings sampling algorithm for Hybrid Monte Carlo\cite{pymc3hybridmontecarlo} as implemented in the PyMC3 package (v 3.0) for Python3.\cite{pymc3} See the \SI~for complete model specification. The model parameters were initialized uniformly and trained for 3000 steps. In all cases, leave-one-out cross-validation was performed via the PyMC3 built-in implementation of Vehtari \textit{et al.} \cite{PYMC3_CROSS_VALIDATION}

The motif model was trained using Gibbs sampling with constrained, per-coordinate infinite horizon stochastic gradient descent using L1 regularization with a squared-difference loss function. An analysis of this method can be found in Mcmahan and Streeter. \cite{SGDInfiniteHorizon} A mathematical description follows.
Given a $D$-dimensional vector of amino acid distributions, let $\boldsymbol{\bar{n}}^{(t)} = N \boldsymbol{X}^{(t)}$ be the vector of  \textit{expected} counts from each distribution at timestep $t$, with $N$ the total number of observations, and let $\boldsymbol{m}^{(t)}$ be the \textit{observed} counts given by the observation (draw) made at step $t$, with regularization term $||\boldsymbol{x}||_1$, and let $\boldsymbol{\nu}$ represent some uniformly random noise vector. We define the loss function $\boldsymbol{L}$, the learning rate $\eta ^{(t)}$, and the regularization term $||\boldsymbol{x}||_1$ as:
\begin{equation}
\label{eq:gibbsdefs}
\begin{split}
\boldsymbol{L} = & \left( \boldsymbol{\bar{n}}^{(t)} - \boldsymbol{m}^{(t)} \right)^2 + ||\boldsymbol{x}||_1\text{,}\\
\eta^{(t)}  = & \frac{1}{\sqrt{\sum\limits_{t=0}^{t-1}\left(  \frac{\partial{\boldsymbol{L}}}{\partial{\boldsymbol{X}^{(t)}}} \right)^2}}\text{, and}\\
& ||\boldsymbol{x}||_1 = \lambda \sum\limits_{i=1}^D |\boldsymbol{\bar{n}}^{t}_i|\text{.}\\
\end{split}
\end{equation}

 Using the definitions in Equation~\ref{eq:gibbsdefs}, and given an initial distribution vector $\boldsymbol{X}^{(0)}$, the update to $\boldsymbol{X}^{(t)}$ at timestep $t$ is given as
 \begin{equation}
 \label{eq:gibbsupdate}
 \begin{split}
     \boldsymbol{X}^{(t+1)} & = \boldsymbol{X}^{(t)} - \eta ^{(t)} \frac{\partial{\boldsymbol{L}}}{\partial{\boldsymbol{X}^{(t)}}} + \boldsymbol{\nu}\text{, with}\\
     \frac{\partial{\boldsymbol{L}}}{\partial{\boldsymbol{X}^{(t)}}} & = 2N \left( \boldsymbol{\bar{n}}^{(t)} - \boldsymbol{m}^{(t)} \right) + \lambda
 \end{split}
 \end{equation}

In general, $L1$ regularization is defined as $|| \boldsymbol{x} ||_1 = \lambda \sum\limits_{i=1}^n |y_i - f(x_i)|$, where the $y_i$ term refers to the ``target value,'' and $\lambda$ is an adjustable parameter. In our case, $y_i = 0 ~ \forall i$. This indicates a low belief in any value above zero for our motif distributions, and induces sparsity in the final distributions. The $\nu$ term is a stochastic noise term that uniformly adds an observation at random at each update step. This helps overcome overfitting by exploring more of the sample space.

Typical cross-validation of this motif model is not sufficient to evaluate all aspects of its performance. Its purpose is not only classification, but also identification of motifs among peptides, which are relatively rare.  We evaluate prediction accuracy via withheld testing data from the APD dataset, but it is important to also validate the intended motif-capturing behavior of the model separately. However, because the entries in the APD do not have labeled motifs, and are not guaranteed to all contain motifs, it is impossible to validate this model's ability to capture motif information from these positive cases. 
Thus, to evaluate the ability to identify motifs, trial runs were performed with arbitrarily constructed, small datasets with imposed motifs. One or more fixed motifs (e.g. QAFR, IEKG, etc.)  were selected, and background members consisting of uniformly-distributed amino acids were appended and prepended to the motifs randomly. For example, test peptides containing the QAFR motif might be ARQAFROI, or IQFARGMO. During training, these datasets had a random 20\% withheld as testing data. The artificially-constructed motifs were captured accurately by the model with as few as 500 iterations over the data set. Figures S5-S8 show the fitted motif distributions for the data with the imposed motif ARND.

Combining the two models requires no additional training step because the two halves of the combined model were trained previously. The trained distributions from the two halves of the model were used to evaluate likelihoods for the positive and negative datasets used previously. These likelihoods are normalized by dividing the likelihoods produced by each half of the model by the highest likelihood in that half. Then, weights $W$ ranging from $0$ to $1$  were assigned to the motif model, with $1-W$ being assigned to the QSPR model. The sum of these two weighted likelihoods for a given peptide is the likelihood produced by the combined model for that peptide. For each weight, a receiver operating characteristic (ROC) curve is generated. The ROC curve is generated by calculating false positive rate (FPR) and true positive rate (TPR) of classification on the {\color{black}training} data as the cutoff value for likelihood is varied between $0$ and $100\%$ of the maximum likelihood produced by the model. A data point that scores a likelihood above the cutoff value indicates a positive (i.e., the model predicts it to be antimicrobial). The FPR is the fraction of such points from the non-antimicrobial (negative) testing set, and the TPR is the fraction from the antimicrobial (positive) testing set.  The accuracy values displayed in Figure~\ref{fig:combinedmodel}b are the accuracy {\color{black}calculated on the withheld testing dataset} at the optimal cutoff for the ROC curve at each weighting on the $x$-axis.

\section{Results: Bayesian Network Models that Predict Activtiy}

\begin{figure}

    \begin{subfigure}{0.45\linewidth}
    \centering
    \includegraphics[width=.8\linewidth]{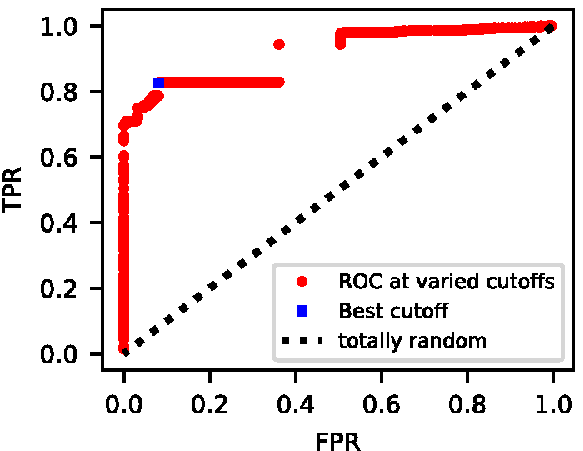}
    \subcaption{The ROC Curve for the 3-kernel Gaussian mixture QSPR model.}
    \end{subfigure}
    \begin{subfigure}{0.45\linewidth}
    \centering
    \includegraphics[width=.8\linewidth]{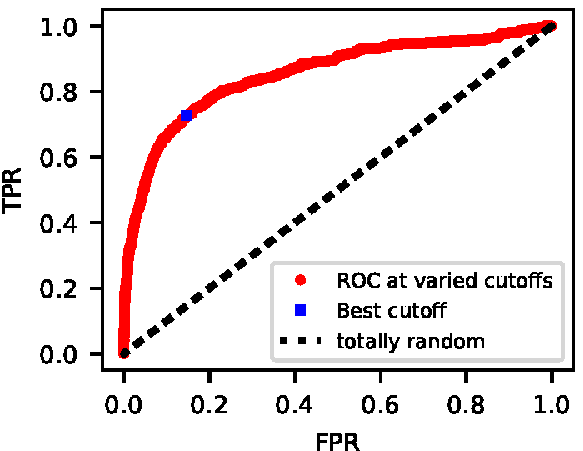}
    \subcaption{The ROC Curve for the motif model with 8 possible motifs of length 3.}
    \end{subfigure}
    
    \caption{Example receiver operating characteristic (ROC) curves for the two model types. TPR is the true positive rate, and FPR is the false positive rate. The best cutoff was defined as the point which minimized $\sqrt{( 2(FPR)^2 +   (1-TPR)^2 )}$. This objective function was chosen to put an emphasis on a lower FPR.}
    
    \label{fig:exampleroc}
    
\end{figure}

We have created three increasingly sophisticated models for predicting activity. Figure~\ref{fig:exampleroc} shows example ROC curves for the best parameter sets for the QSPR and motif models. A summary of results using the best parameter sets for the three models are shown in Table~\ref{tbl:qspr2:design}.

{\color{black}Performance of each model was evaluated based on the accuracy produced by the model at the point on the ROC curve (generated using withheld testing data) that minimized the value $\sqrt{(2(FPR)^2 + (1-TPR)^2)}$. This choice was made to emphasize a lower FPR. In the case of the Gaussian mixture model, the kernel number was varied between $2$ and $10$, and accuracy was evaluated for all three of the chosen chemical descriptors with that kernel number. This choice was made to simplify model specification and to limit the number of training sessions. While the same kernel number may not be optimal for each individual descriptor, the model produced sufficient accuracy with this simplification. For the motif model, motif count varied between $k=2$ and $k=10$, and motif length varied between $w=3$ and $w=8$. The ``background-only'' (i.e. $k=w=0$)  model was also evaluated.}

Some QSPR trials with different kernel numbers produced very similar performances. In particular, the 3-kernel and 6-kernel trials both produced accuracy above 80\%. Thus, it was necessary to decide which trial's distributions to use for testing. The 3-kernel data was used after considering the distributions depicted in Figure S9. The 3-kernel and 6-kernel QSPR models performed nearly identically in terms of ROC and prediction accuracy (Figure S9a and S9b), but comparing Figure S9c and S9d shows us that the 3-kernel model's distribution has a shape that is more indicative of the underlying structure of the raw data.

The motif model had optimal performance with $k=8$ and $w=3$, and the resultant ROC curve is shown in Figure~\ref{fig:exampleroc}b. A heatmap of accuracies with different $k$ and $w$ values is shown in Figure~\ref{fig:qspr2:dgm}c.
Surprisingly, the model's optimal performance with motifs is of identical accuracy with $0$ motifs (background only), with both the $k=8$, $w=3$ and $k=0$, $w=0$ models having an accuracy of $79\%$ with optimal cutoff. The ROC curve of the background-only motif model is shown in Figure S10. This may indicate that motifs are not important for antimicrobial activity, but merely the overall distribution of amino acids present, or it could indicate that motifs play a more complicated role in antimicrobial behavior than this model is able to capture.

The parameters that gave the best performance for the two individual models were also used for the combined model. The weights assigned to each half of the model were varied continuously from $0$ to $1$ to determine the optimal hyperparameter. Likelihoods from the two models were re-weighted by dividing all likelihoods for one model type by the highest likelihood produced by that model, to achieve comparable magnitudes from the two models. These results are shown in Figure~\ref{fig:combinedmodel}b. With its best performance, the combined model outperformed both of the individual models, with a weight of $21\%$ for the motif part of the model and $79\%$ for the QSPR part producing a classification accuracy of $94\%$.

\begin{figure}
\centering

\includegraphics[width=8cm]{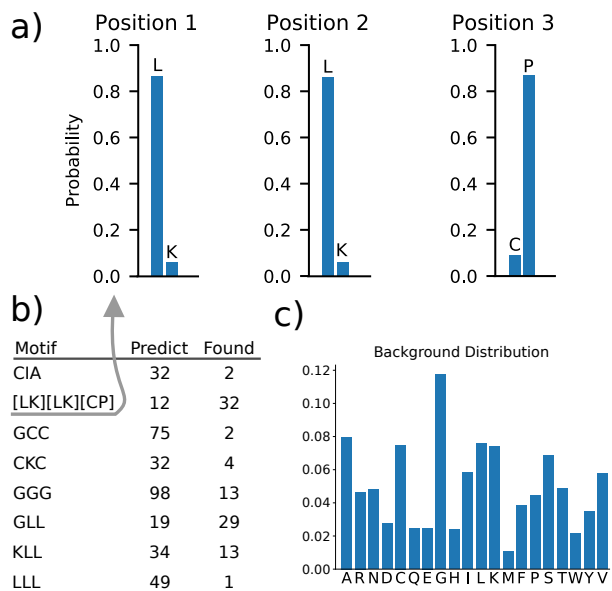}
\caption[Motif model learned parameters]{Panel a shows the probability that a given amino acid appears in each of positions 1 through 3 in the motif ``[LK][LK][CP]''. Due to the sparsity
  from regularization, the majority of the motifs predicted by the model are sparse (only one amino acid with non-negligible probability for that motif position). Panel b is the list of motifs predicted by the
  model. The ``Predict'' column is the number of sequences which are
  more likely to contain the corresponding motif than any other motif. The
  ``Found'' column is the number of sequences that actually contain the
  motif. Panel c is the background distribution of amino acids from
  the motif model. The $y$-axis is probability of observing a randomly chosen amino acid from a random peptide in this dataset.}
\label{fig:qspr2:motifs}

\end{figure}

The ability to interpret the model may be seen in Figure
\ref{fig:qspr2:motifs}. Figure \ref{fig:qspr2:motifs}a shows the
probability distribution from the second motif from the $k=8$, $w=3$ model. The regularization
operates as expected and the motifs are sparse; only one or two amino
acids have non-negligible probability for a given position. Figure \ref{fig:qspr2:motifs}b shows
all the motifs predicted by the model. The ``Predict'' column shows the count of peptides for which the given motif had the highest likelihood of appearing. This should be interpreted as the ``best match'' motif for a given peptide. It does not imply that the peptide contains that motif, but only that the shown motif gives the highest likelihood for that peptide among all motifs predicted by the model. The next column contains the
number sequences which actually contain each motif, obtained by exhaustive analysis. We see the model
has correctly assigned each motif to the corresponding sequence based
on the close match between the predict and found columns. There are relatively few examples of the motifs
discovered by the model, but it does capture some common ones. For example, exhaustive analysis shows that GLL is the third most common 3-letter motif in the APD dataset, and it is captured by the model, as shown in Figure~\ref{fig:qspr2:motifs}b. GGG is also among the 10 most common motifs found from brute-force analysis, and is also captured by the model. However, the model did not capture all the most common motifs. Finally, the background distribution is shown in Figure
\ref{fig:qspr2:motifs}c. This may be considered the amino acid
composition of the APD without the motifs observed in sequences. It is not uniform, and different from the Human database
\cite{White2012Decoding} and as mentioned above, contributes
significantly to the performance of the classifier. This is not
unexpected since amino acid composition is a well-used descriptor for
analyzing peptides and proteins.\cite{aaComposition} 
Furthermore, the identical accuracy of the best-case motif model and ``background-only'' model, as well as the failure of the model to capture all the most common motifs, indicates that either motifs are unimportant in the antimicrobial activity of peptides, or the model is insufficient to capture the important aspects of peptide motifs.

In order to compare our model with a traditional machine learning method, we also evaluated the performance of a linear support vector machine (SVM) on the chemical descriptors used in the QSPR model. The datasets used were the same. We utilized the builtin SVM method of the scikit-learn Python 3 package,\cite{scikit-learn} with $3000$ training steps. After convergence, the SVM had an accuracy of $84\%$ with its optimal cutoff, which is not as good as the QSPR, or QSPR + motif models, which were $87\%$ and $94\%$, respectively.

\section{Discussion: Identifying Multi-Functional Peptides}

As shown above, it is possible to construct accurate models that can predict peptide
activity. These tools can be further used to find
peptides which have multiple activities or multiple functions. As
stated in the introduction, the Human dataset contains peptides which
are likely antifouling. To find a peptide that is both antimicrobial
and antifouling, we can identify a peptide from the APD that scores
as active according to a model trained on the Human dataset. The
opposite procedure is possible using the models trained
above, where we find a human protein fragment that is likely
antimicrobial. However, there is no experimental evidence that such a
fragment is antifouling. Choosing a peptide from the APD that is
human-like will guarantee at a minimum that it has antimicrobial
activity, and that the model predicts it is similar in character to sequences found on the surfaces of human proteins. There
is no evidence showing that motifs are relevant for antifouling, but past research has shown that strong net neutral partial charges, hydrophilicity, and low self-interaction are,\cite{White2013Standardizing,white2014stealth} so
a QSPR model was used.

\begin{table*}
\centering

\begin{tabular}{lcccr}
Model & FPR & TPR & Accuracy & \color{black}MCC\\
\hline
QSPR  & 8.1\% & 83\%  & 87\% & \color{black}0.75\\
Motif & 17\% & 75\% & 79\% & \color{black}0.58\\
QSPR + Motif & 3.4\% & 90\% & 94\%  & \color{black}0.87\\
\end{tabular}
\centering
\caption{Summary of models that best predicted antimicrobial
  activity. The QSPR model is depicted in Figure~\ref{fig:qspr2:classifier}, the motif model in Figure~\ref{fig:qspr2:dgm}, and the QSPR + motif in Figure~\ref{fig:combinedmodel}a. FPR and TPR are false positive and true positive rates of classification, respectively. \color{black}MCC is the Matthews correlation coefficient. \label{tbl:qspr2:design}}

\end{table*}

The QSPR model described above was used. The model was
fit to the Human dataset with the same procedure as was used on the APD. The 8-kernel QSPR model performed the best for this dataset, with an accuracy of $67$\%. Overall, the model performance on the Human dataset was worse than on the APD, as can be seen by the ROC curve in Figure S11. This is likely due to the multimodality in the Human dataset, as well as its broad definition of activity (being present on the surface of a human protein). However, the model still achieved a low FPR, as desired. This shows that the model works regardless of the dataset it is trained against.

After analyzing the descriptors of the APD against the optimal cutoff for the 8-cluster QSPR model trained on the Human dataset, approximately $30$\% of the APD peptides were found to be human-like. A peptide was designated as ``human-like'' if it scored above the cutoff used to produce the optimal accuracy on the ROC curve of the QSPR model fitted to the Human data. This low percentage shows AMPs are generally different from human proteins
surfaces, which are thought to be optimized for minimal nonspecific
interactions.\cite{White2012Decoding,  White2012Role}

After omitting sequences less than 30 amino acids long, the most
human-like AMPs are WKSESLCTPGCVTGALQTCFLQTLTCNCKISK (APD number AP00206) and ITSISLCTPGCKTGALMGCNMKTATCHCSIHVSK (APD number AP00205). The first is subtilin, an antibiotic produced by the bacterium and model organism \textit{Bacillus subtilis}.\cite{subtilin,apd} The second, nisin A, is produced by the bacterium \textit{Lactococcus lactis}, a species used in the production of cheeses.\cite{nisinA,apd} They are similar to sequences from the ``Human'' dataset due to
their low number of non-polar groups, high number of charged groups,
and slightly negative net charge. Due to the connection between low
protein adsorption and human protein surfaces,\cite{White2012Decoding}
these two sequences may be good candidates for stable (non-fouling) antimicrobial surface
coatings. Both of these sequences underwent PTM before antimicrobial behavior was observed, yet the model was still able to predict their antimicrobial nature \textit{without this information}. This demonstrates that it is possible to predict antimicrobial potential for a given sequence without knowledge of PTMs. Thus, this model has the advantage of needing little information while still providing high classification accuracy, but it also has the disadvantage of being unable to predict whether PTMs are necessary for activity. Although this model cannot indicate whether PTM will be necessary for a given peptide, it could still be used to screen or evaluate candidate sequences for PTM experiments.

\section{Conclusions}
The application of Bayesian network models to QSPR {\color{black}peptide modeling techniques} has been
introduced utilizing open-source statistical modeling
software.\cite{pymc3} These models are flexible and may encode
sophisticated chemical knowledge, as seen from the motif model
presented. This flexibility also allows models to be constructed with
easy to interpret parameters, as demonstrated by the
motif and combined QSPR + motif models, where regularization forced each motif position to only
contain one amino acid, as opposed to previous models where motif
positions have non-negligible probability assigned to each of the 20 amino acids
.\cite{White2013Standardizing, Bailey1994MEME} These models
show good classification performance with a maximum of $94\%$ at
predicting whether a peptide is active against gram-positive
bacteria, given only the sequence of amino acids and their chemical descriptors. This is as good as more opaque and complex strategies
such as multilayer artificial neural networks\cite{bestAPmodel} and N-gram representation random forest modeling,\cite{RandomForestAMPs} and better than a linear SVM, with the advantage of chemically meaningful interpretations.
Additionally, these models were used to identify potentially
multifunctional peptides that are both antifouling and
antimicrobial. Finally, due to the identical performance of the best-case motif model and the ``background-only'' model, we can conclude that either motifs are unimportant  to the antimicrobial activity of peptides, or their importance is more complicated than this model is able to capture.
Bayesian network models provide a significant advance in the type
of peptide activity modeling that can be done, and the ease in which such
models can be constructed and combined.

\section{Acknowledgement}
This work was supported by the Office of Naval Research
(N00014-10-1-0600) and the National Science Foundation (CBET-0854298).

\bibliography{qspr3} 
\bibliographystyle{plainnat}

\newpage

\end{document}


\maketitle
\vspace{0.1cm}

\begin{centering}
\small
$^\dagger$ Department of Chemical Engineering\\
University of Rochester\\
206 Gavett Hall\\
Rochester, NY 14627, USA\\\vspace{0.1cm}
Email: andrew.white@rochester.edu\\\vspace{0.1cm}
$^\ddagger$ Department of Chemical Engineering\\
University of Washington\\
Box 351750\\
Seattle, WA 98195, USA\\\vspace{0.1cm}

$^*$ To whom correspondence should be addressed

\end{centering}

\section*{Converting Descriptors into Ranks}

In order to use a structural descriptor in a Bayesian network model, it must
be converted into a form that may be described by a probability
distribution. The probability that a descriptor $f(\cdot)$ equals a value $x$ in a
compound $c$ is given by:
\begin{equation}
\label{eq:prob}
\Pr(f(c) = x) = \frac{1}{Z}\sum_i w_i \Ind{f(c)}{x},\;Z = \sum_i w_i
\end{equation}
where $Z$ is the partition coefficient, $w_i$ is the un-normalized
probability (weights) of observing the $i$th compound in the chemical
space, and $\delta$ is an indicator or delta-function. The weighting
parameters may be adjusted, for example, to account for synthetic
difficulty, recognizing the fact that the experimentally active
compounds are likely chosen with bias. In the case of peptide
libraries, the weights are unity because peptide libraries have little
to no synthetic bias. The partition coefficient for a peptide library
is $A^l$, where $A$ is the size of the alphabet (generally 20 for
amino acids) and $l$ is the length of the amino acid sequence.

Constructing probability distributions for group-wise additive
descriptors follows two approaches. When the chemical space is small
($l \leq 3$ for peptides), all descriptor values may be enumerated to
create a probability distribution. When the chemical space is large
($l > 3$ for peptides), the probability distribution may be
approximated as a sum of $l$ normal distributions. In the case of
peptide libraries, $l$ is the length and the normal distributions are
identical. In the case of combinatorial organic libraries, $l$ is the
number of positions that may be exchanged and the normal distributions
may not be identical between positions. The approximation is accurate
provided the number of $0$'s is low (e.g., the number of sulfur atoms
in the peptides will not fit into this approximation). Examples of
this approximation may be seen in Figures \ref{fig:MWFitting}-\ref{fig:NetChargeFitting}. The mean of the
normal distributions is the mean ($\mu$) of the descriptor calculated
on the combinatorial components (e.g, amino acids) and the variance
($\sigma^2$) is calculated likewise. The sum of the $l$ normal
distributions will have a mean of $l\mu$ and variance $l\sigma^2$. For
non-group-wise distributions, the probability distribution may be
estimated by sampling from the combinatorial library where the
sampling is done according to the weights $w_i$.

\begin{figure}

\centering
\includegraphics[width=8cm]{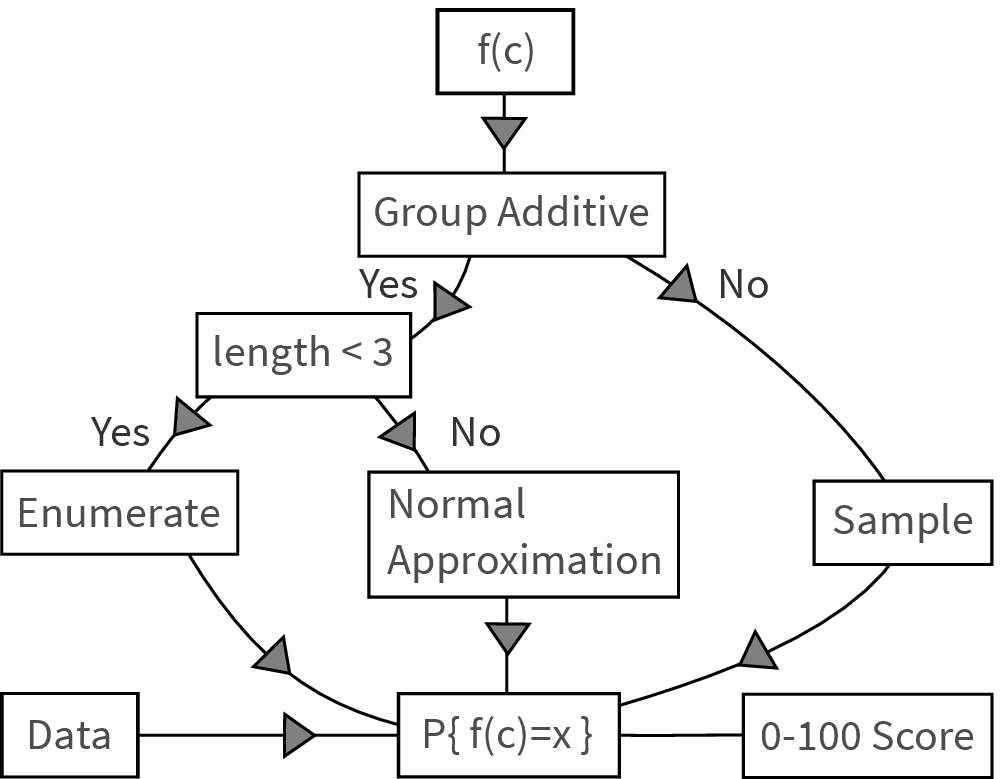}
\caption[Flowchart for converting QSAR descriptor into rank]{A
  flowchart for converting a descriptor, $f(c)$, into a rank from
  0--100 that both removes biases from the chemical space of the
  library and normalizes it for use in Bayesian network models.}
\label{fig:qspr2:process}

\end{figure}

\begin{figure*}
\centering
\includegraphics[width=8cm]{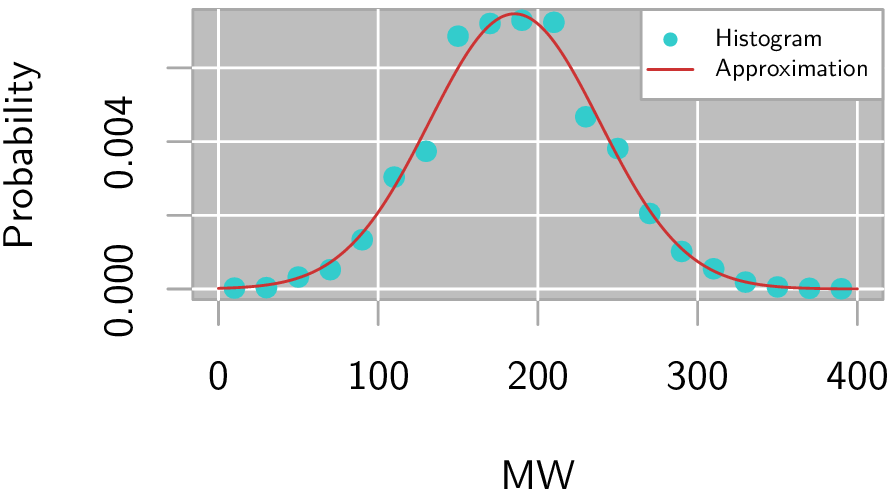}
\caption[Comparison of exact molecular weight and approximation
  described in main text.]{ The dataset is all combinations of the 20 amino
  acids with length 3. The approximation is the sum of three identical
  normal distributions parameterized to the molecular weight of the 20
  amino acids. The approximation works well, even at this low length.}
  \label{fig:MWFitting}
\end{figure*}

\begin{figure*}
\centering
\includegraphics[width=8cm]{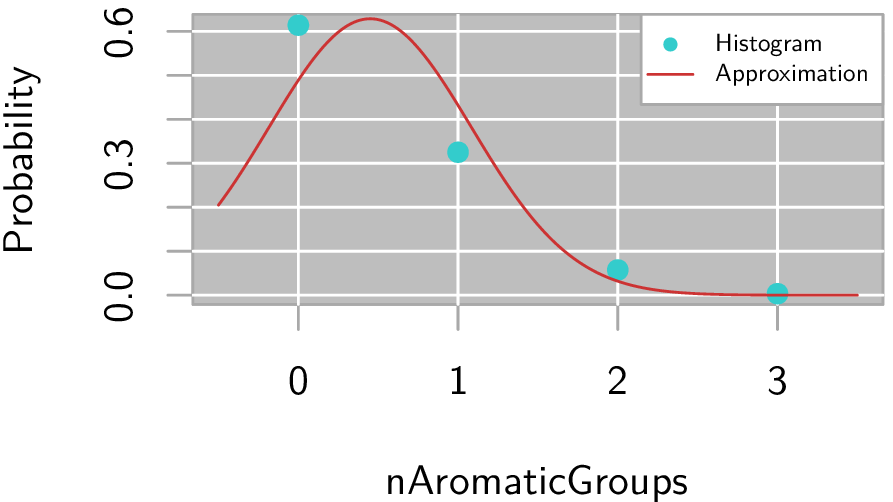}

\caption[Comparison of exact count of aromatic groups and
  approximation described in main text.]{ The dataset is all
  combinations of the 20 amino acids with length 3. The approximation
  is the sum of three identical normal distributions parameterized to
  an aromatic indicator function on the 20 amino acids (1 for
  aromatic, 0 for non-aromatic). The approximation doesn't works well,
  due to the high number of zero values, as mentioned in the text.}
  \label{fig:nAromaticGroupsFitting}
\end{figure*}

\begin{figure*}
\centering
\includegraphics[width=8cm]{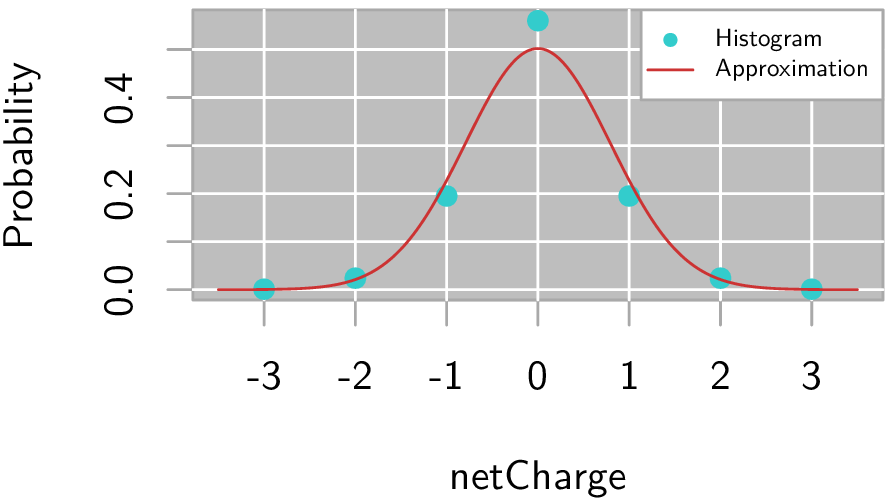}
\caption[Comparison of exact net charge and approximation described in
  main text.]{ The dataset is all combinations of the 20 amino acids
  with length 3. The approximation is the sum of three identical
  normal distributions parameterized to the charge of the 20 amino
  acids. The approximation works well, even at this low length.}
  \label{fig:NetChargeFitting}
\end{figure*}

Once the probability distribution over the chemical space of the
library is calculated, descriptors for the active/training compounds
are transformed to incorporate information about this probability
distribution. This is done by converting the descriptor into a rank
between 0 and 100, where the rank of the descriptor
relative to the chemical space. The ranks come from quantiling the
descriptors calculated over the chemical
space\cite{Hyndman1996Sample}. For example, quantiling the number of
charged groups over a chemical space with 4 quantiles could yield that
the bottom 25\% are between 0--5 charge groups, the 25--50\% are
between 5--6 charge groups, 50--75\% are between 6--7 and the top 25\%
are 7--15. Using this distribution, an active compound with 3 charged
groups would be given a rank of 1, because it is in the first
quantile. An active compound with 7 charge groups has a rank of 3 and
an active compound with 13 charges would also has a rank of 3
. Notice how the unevenness of the original distribution is removed
and the ranks correlate to the ranking over the chemical
space. The entire transformation process is depicted in
Figure \ref{fig:qspr2:process}.

This descriptor transformation has three benefits. First, it is
immediately obvious if a descriptor is at an extreme value. Second,
when examining multiple descriptors, their range corresponds exactly
to their span of the entire chemical space. Thus, if a descriptor
range is 5--95, it is not significant. If it is within the range of
20--25, then the descriptors occupy a range that only 5\% of the
chemical space of the library occupies. Third, the effect of length on
the peptide descriptors may be removed by only comparing descriptors
against uniform length probability distributions. For example, if
there are sequences from lengths 3--10 in a library, the descriptors
may be calculated relative only to sequences of the same length. Then
a rank of 5 is interpreted as in the bottom 5\% relative to sequences
of the same length. If this is not desired, only the probability
distribution on the longest 2 lengths need to be calculated since that
corresponds to 99.75\% $(1 - 1/ 20^2)$ of the possible values.

\section*{Model specifications}
The grphical models are specified in the GitHub repository found at \url{https://github.com/RainierBarrett/pymc3_qspr}. This software is made freely available under the GNU General Public License.

\section{Motif Model Verification}
\begin{figure}
    \centering
    \includegraphics[width=0.75\textwidth]{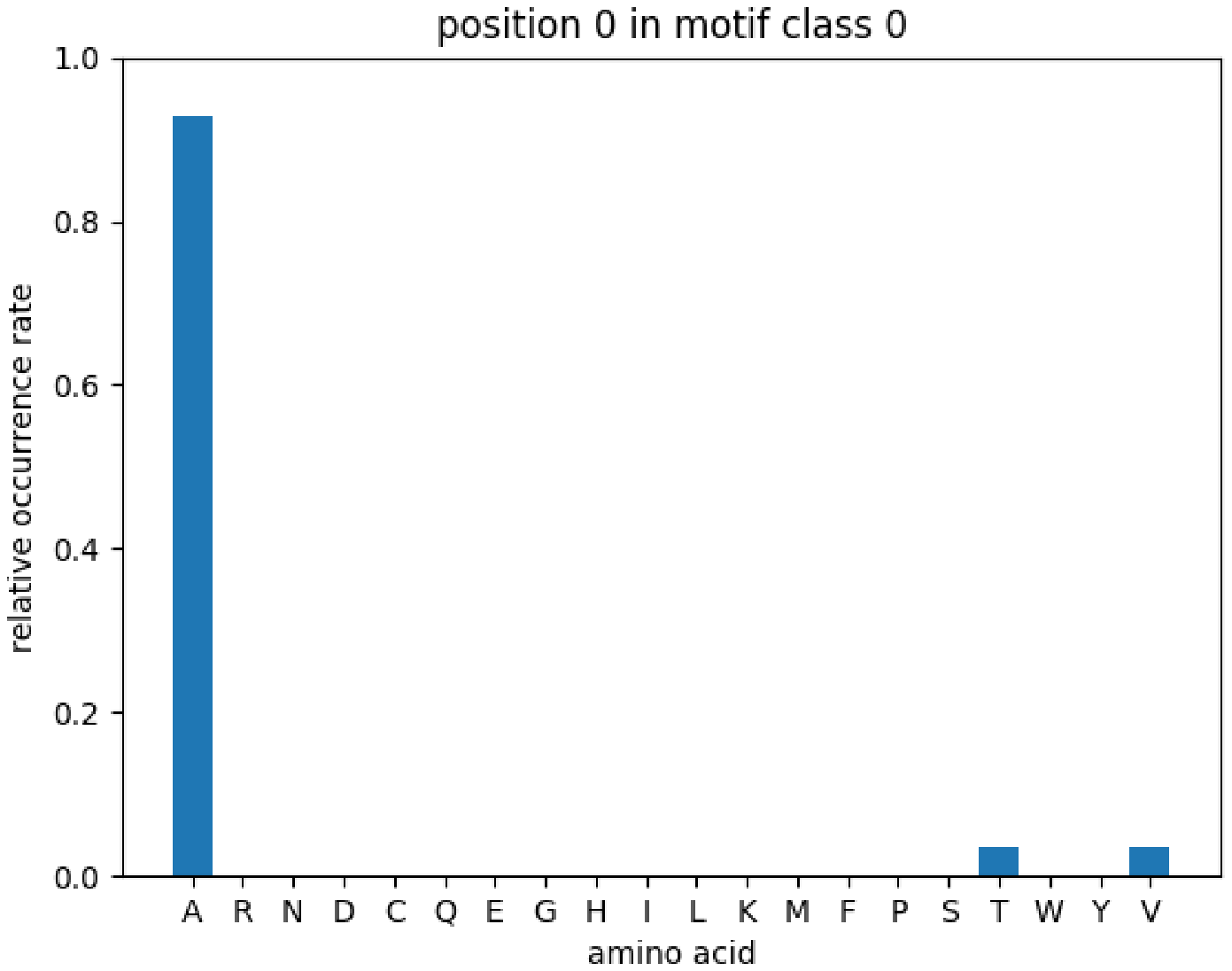}
    \caption{Predicted probabilities for amino acid ocurrence in the first position in a single motif class model.}
    \label{fig:singlemotif1}
\end{figure}

\begin{figure}
    \centering
    \includegraphics[width=0.75\textwidth]{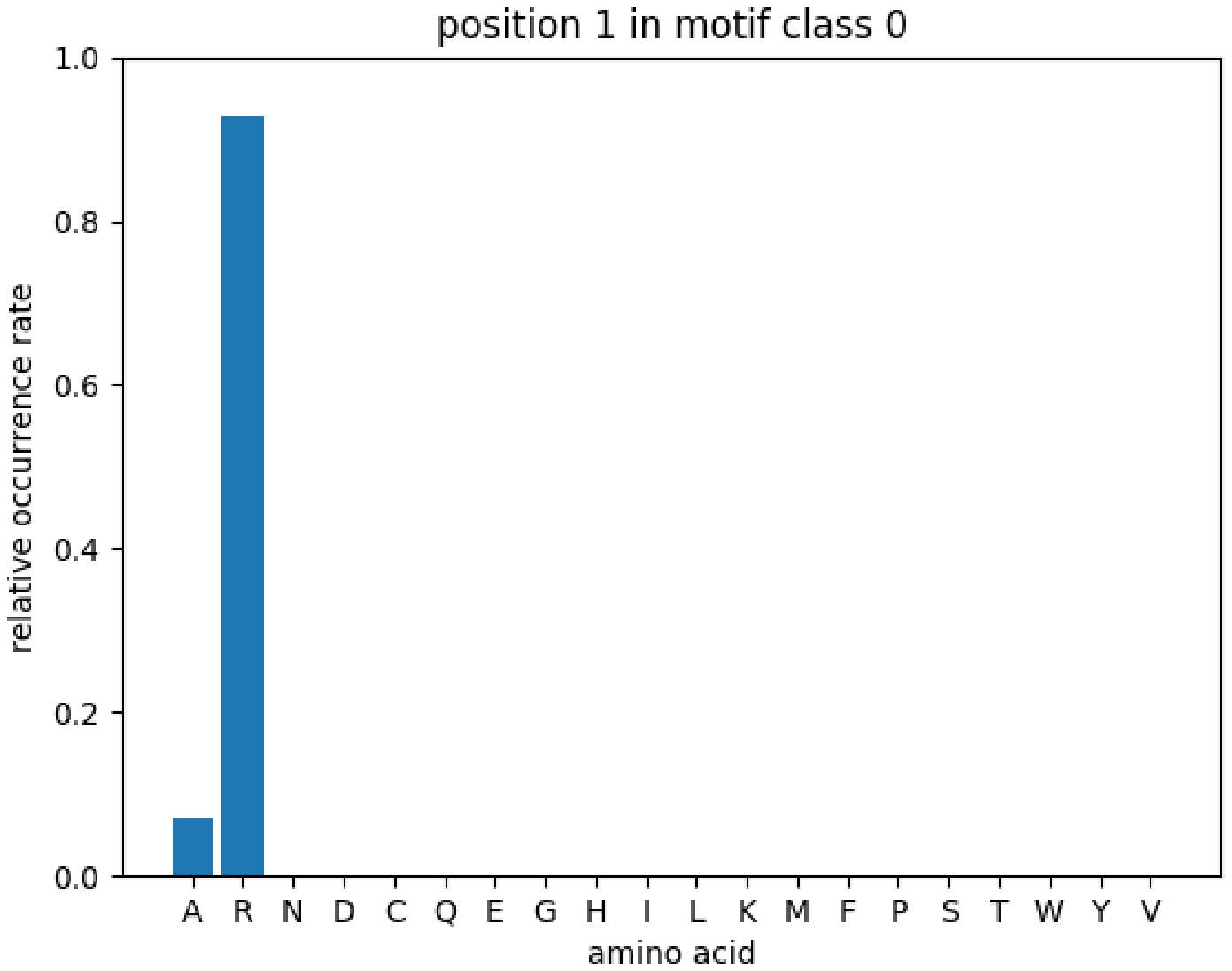}
    \caption{Predicted probabilities for amino acid ocurrence in the second position in a single motif class model.}
    \label{fig:singlemotif2}
\end{figure}

\begin{figure}
    \centering
    \includegraphics[width=0.75\textwidth]{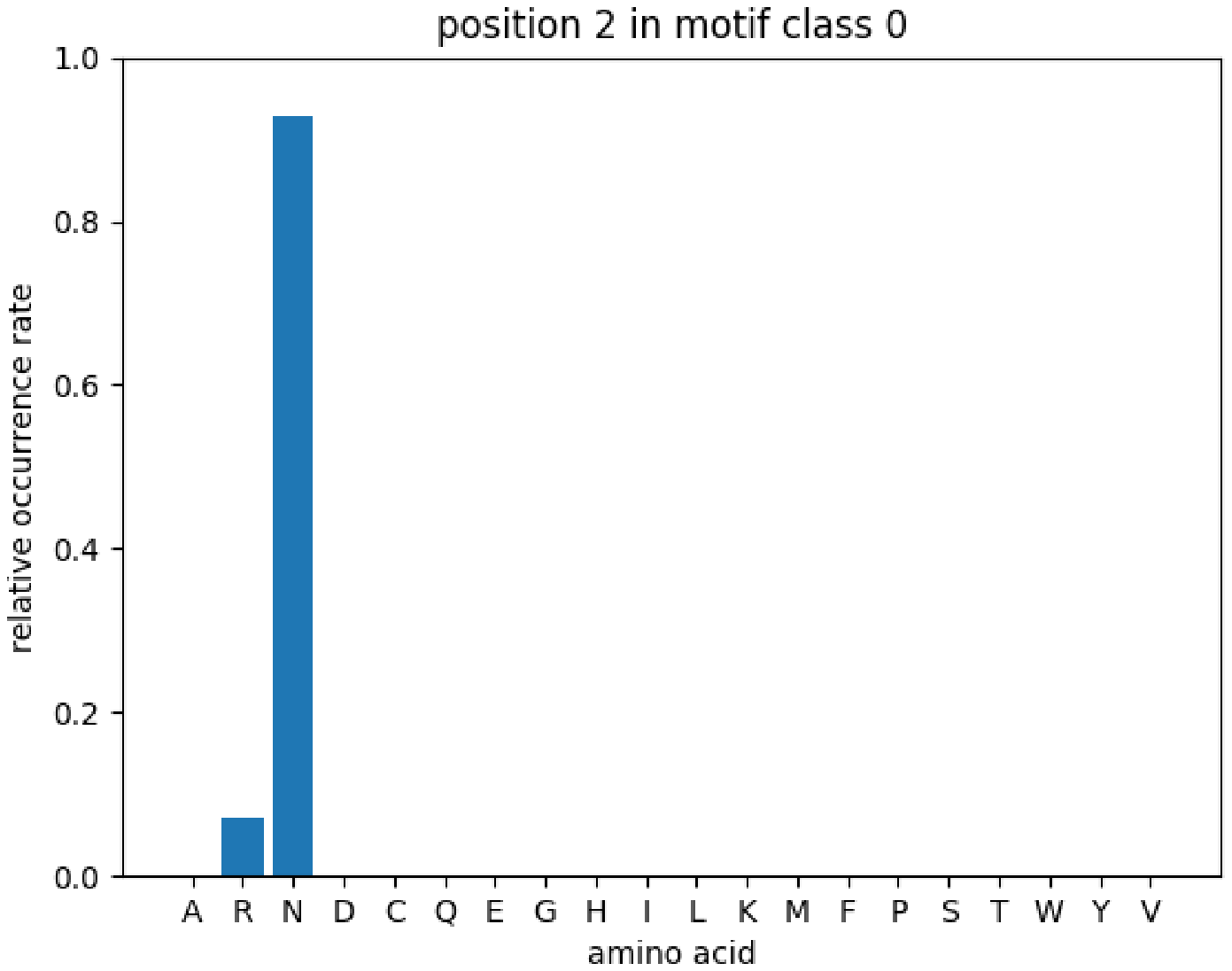}
    \caption{Predicted probabilities for amino acid ocurrence in the third position in a single motif class model.}
    \label{fig:singlemotif3}
\end{figure}

\begin{figure}
    \centering
    \includegraphics[width=0.75\textwidth]{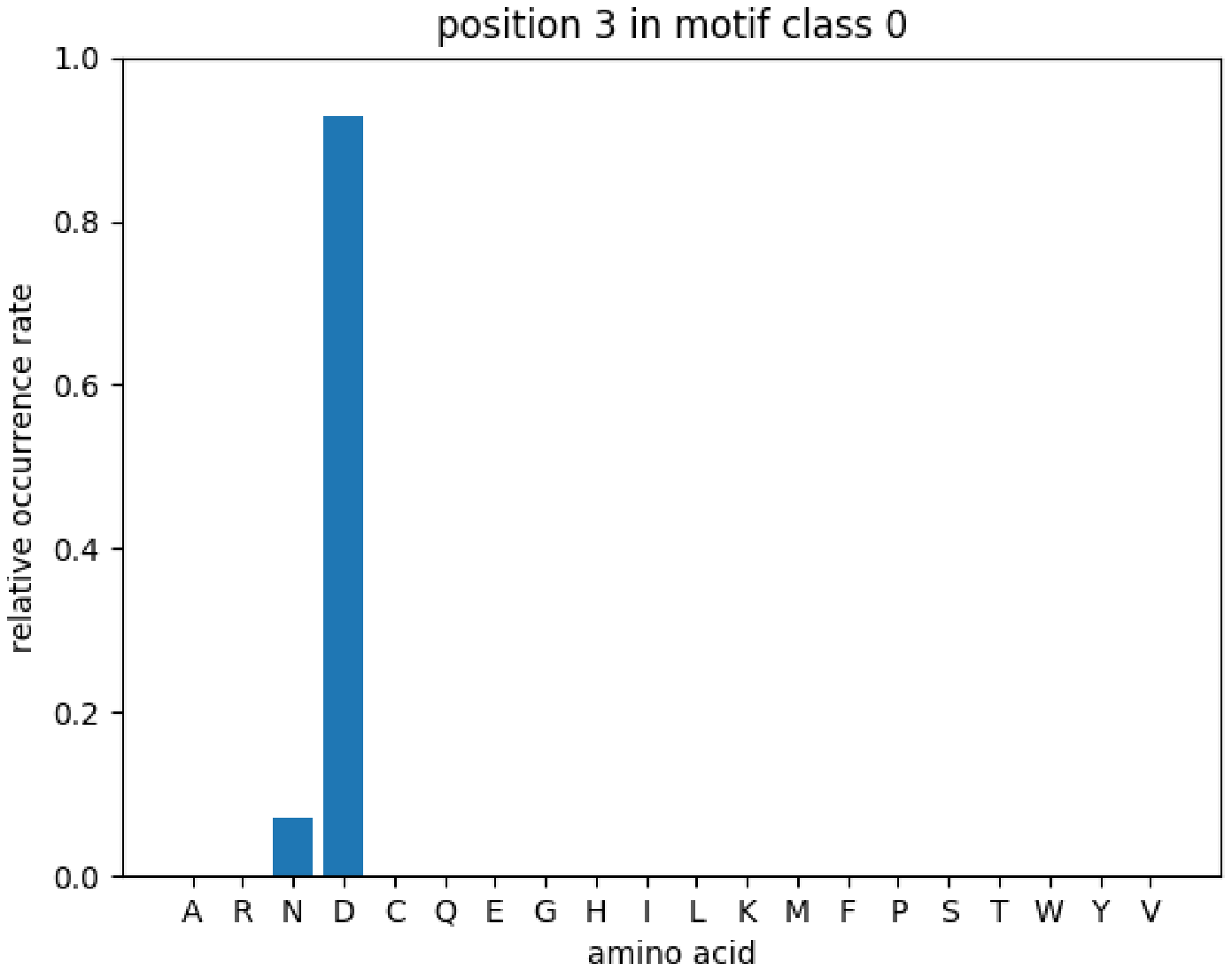}
    \caption{Predicted probabilities for amino acid ocurrence in the fourth position in a single motif class model.}
    \label{fig:singlemotif4}
\end{figure}

Figures \ref{fig:singlemotif1}-\ref{fig:singlemotif4} are the plots of the motif class distribution trained on the artificial dataset with the imposed motif ``ARND''. As described in the main text, the dataset of peptides with imposed motifs was artificially constructed by adding uniformly-distributed amino acids before and/or after the imposed motif randomly. Notice the sparseness -- the model clearly distinguishes which amino acid is most likely to be in which motif position, with no knowledge of what the motif will be or where it will occur. The model accurately captures the motifs in this simple case. These figures were produced by fitting the motif model with 1000 training steps with the same training method as described in the main text.

\begin{figure}

    \begin{subfigure}{0.45\linewidth}
    \centering
    \includegraphics[width=.7\linewidth]{3_cluster_ROC.eps}
    \subcaption{The ROC Curve for the 3-kernel Gaussian mixture QSPR model.}
    \end{subfigure}
    \begin{subfigure}{0.45\linewidth}
    \centering
    \includegraphics[width=.7\linewidth]{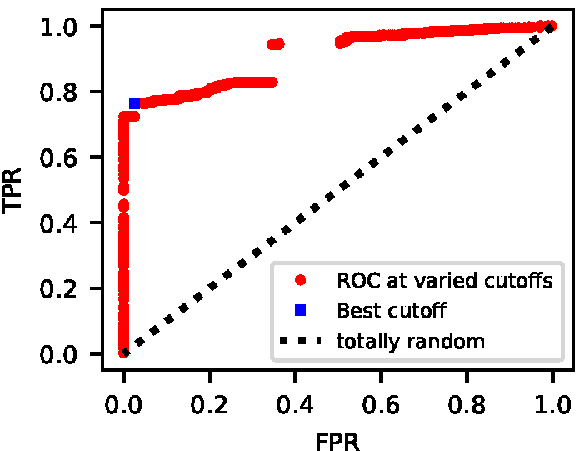}
    \subcaption{The ROC Curve for the 6-kernel Gaussian mixture QSPR model.}
    \end{subfigure}

    \begin{subfigure}{0.45\linewidth}
    \centering
    \includegraphics[width=.8\linewidth]{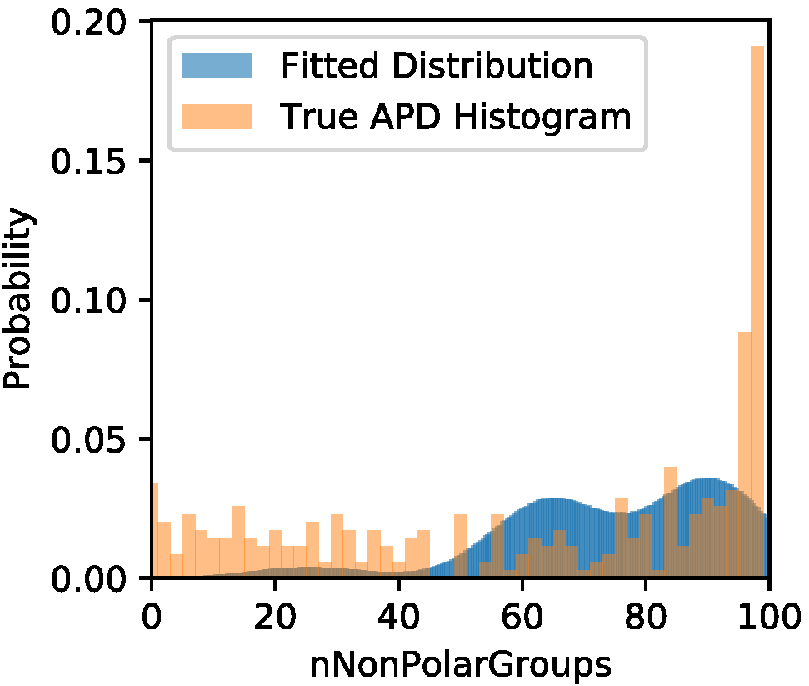}
    \caption{The fitted distribution for number of non-polar groups score in the 3-kernel QSPR model compared with the histogram of the raw data.}
    \end{subfigure}
    \begin{subfigure}{0.45\linewidth}
    \centering
    \includegraphics[width=.8\linewidth]{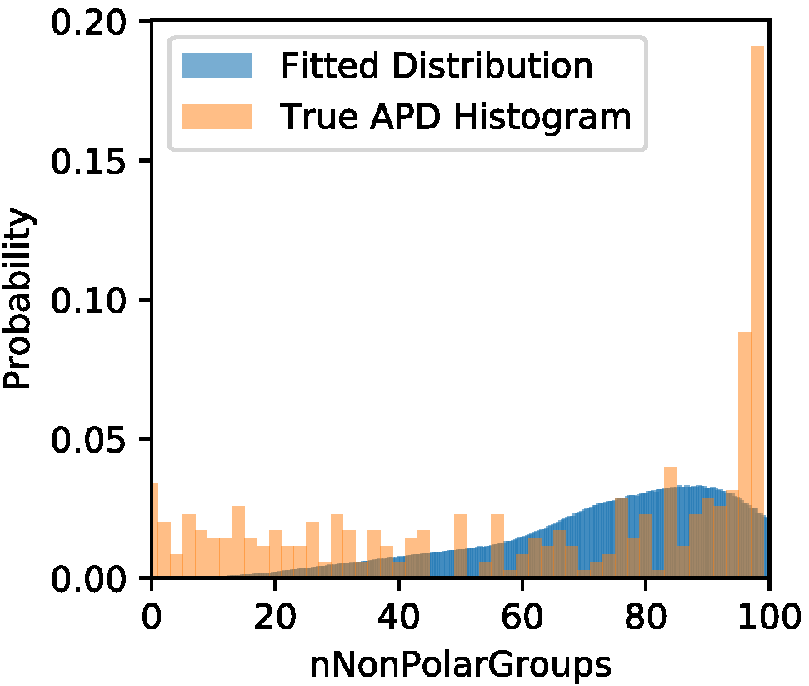}
    \caption{The fitted distribution for number of non-polar groups score in the 6-kernel QSPR model compared with the histogram of the raw data.}
    \end{subfigure}
    
    \caption{A comparison of one descriptor's prediciton histogram with two different kernel numbers in the QSPR model, and the corresponding ROC curves. TPR is true positive rate and FPR is false positive rate for predictions on withheld testing data. Though both of these kernel numbers produced similar performance (note ROC curve similarity), the 3-kernel distribution is more indicative of the underlying properties.}
    
    \label{fig:clusterscompare}
    
\end{figure}

\begin{figure}[H]

    \centering
    \includegraphics[width=.5\linewidth]{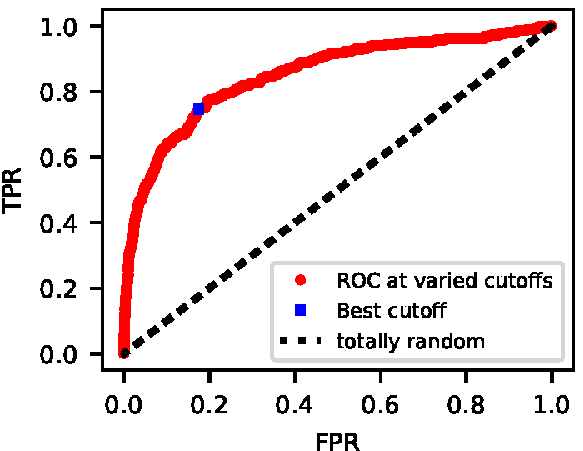}
    
    \caption{The ROC curve generated by the motif model with $k=0$, $w=0$. TPR is true positive rate and FPR is false positive rate for predictions on withheld testing data. The high degree of accuracy with no motif identification may indicate that motifs are not important for antimicrobial activity, or that the model is not equipped to capture the nature of their importance.}
    
    \label{fig:backgroundROC}
    
\end{figure}

\begin{figure}[H]
    \centering
    \includegraphics[width = 3.5 in]{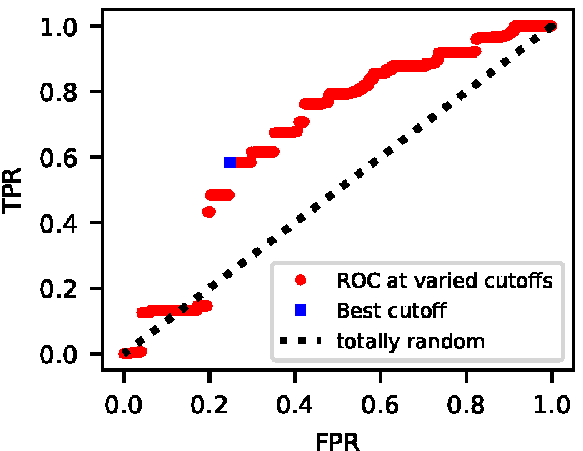}
    \caption{ROC curve of the 8-kernel QSPR model trained on the human dataset. FPR is false positive rate, and TPR is true positive rate for predictions made by this model on  withheld testing data from the human dataset.}
    \label{fig:humanroc}
\end{figure}

Figure~\ref{fig:clusterscompare} shows a comparison between two equally-accurate kernel numbers for the QSPR model. The receiver operating characteristc (ROC) curves and predicted rank distribution for the number of non-polar groups descriptor generated by the model is shown for each case. The Materials and Methods section of the main text details how the ROC curves are generated. We can see that while the ROC curves (and thus, performance) are nearly identical, the three-kernel generated distribution is more representative of the qualitative features of the true histogram of the rankings found from the APD dataset.
Figure~\ref{fig:backgroundROC} displays the ROC curve of the background-only motif model. Note its similarity to the best-case motif model (Figure 6b, main text).
Figure~\ref{fig:humanroc} shows the ROC curve generated by the QSPR model on the human dataset. Note the dissimilarity between this ROC curve and that of the QSPR model trained on the APD (Figure~\ref{fig:clusterscompare}a).

\newpage

\makeatletter
\renewcommand\@biblabel[1]{[#1]}
\makeatother
\bibliography{thesis.bib} 
\bibliographystyle{chem-acs}